\documentclass[preprint,preprintnumbers,amsmath,amssymb]{revtex4}
\usepackage{bm}

\begin{document}

\title{Decoherence of quantum discord in an asymmetric-anisotropy spin system}
\author{Xiang Hao}
\altaffiliation{Corresponding author,Email:110523007@suda.edu.cn}

\author{Chunlan Ma}

\author{Jinqiao Sha}

\affiliation{Department of Physics, School of Mathematics and
Physics, Suzhou University of Science and Technology, Suzhou,
Jiangsu 215011, People's Republic of China}

\begin{abstract}

The decoherence of quantum correlation is investigated in the
Heisenberg spin system with the asymmetric anisotropic interactions.
The quantum entanglement and discord are used to quantify the
quantumness of the correlations. By the analytical and numerical
methods, we find that quantum discord decays asymptotically in time
under the effects of the independent local Markovian reservoirs.
This is markedly different from the sudden change of the
entanglement. Before the disappearance of the entanglement, the
dynamic behaviour of quantum discord is very similar to that of the
entanglement. It is also shown that the discord declines rapidly for
the interacting spin system compared with the case of noninteracting
qubits. At an arbitrary finite temperature, the nonzero thermal
discord can be enhanced by the asymmetric anisotropic interactions
which induce quantum fluctuations.



\end{abstract}

\maketitle

\section{Introduction}

The quantum aspects of correlations in composite systems are a key
issue in quantum information theory \cite{Nielsen00}. Quantum
entanglement is extensively regarded as one useful measure of
nonlocal coherence. The inseparable (entangled) states are
applicable in many parts of quantum information processing, like
quantum teleportation, cryptography, dense coding and quantum
computation \cite{Bennett93,Divin98,Horodecki09}. However, the
recent theoretical and experimental developments have discovered
that some separable (nonentangled) mixed states are useful for
improving performance in some tasks of quantum computers
\cite{Meyer00,Datta08,Lanyon08}. The entanglement is not the only
type of quantum correlation. To generally quantify the quantumness
of correlations contained in bipartite systems, Ollivier and Zurek
have defined a measure known as quantum discord \cite{Ollivier01}.
It is largely accepted that the total correlations contained in
quantum states can be described by quantum mutual information which
is a sum of the quantum discord and classical correlation given by
Henderson and Vedral \cite{Henderson01,Vedral03}. Quantum discord
has received much attention in theoretical studies
\cite{Luo08,Sarandy09,Lutz09,Werlang09,Cui10,Mazhar10,Maziero10}. By
means of the thermodynamics of correlations, quantum discord can
exhibit signatures of quantum phase transitions \cite{Sarandy09}. As
is well known, the composite system loses unavoidably the quantum
coherence due to interacting with the
environments\cite{Rosario08,Leung03}. The dynamics of the discord
were also analyzed in the system of noninteracting qubits coupled to
the environments \cite{Werlang09}. The robustness of quantum discord
is beneficial for the realization of scalable quantum computing in
contrast to the fragility of the entanglement \cite{Yu04}.

Because of the fundamental and practical values, it is necessary to
investigate the dynamic properties of quantum discord in some
concrete systems of interacting spins. In semiconductor quantum dots
and single molecular magnets, the magnetic properties of these
systems were usually studied by the Heisenberg spin models with the
asymmetric anisotropic couplings such as Dzyaloshinskii-Moriya
interactions
\cite{Dzyaloshinskii,Moriya,Dender97,Levitov01,Raedt04}. In this
paper, the Markovian decoherence of the quantum discord is obtained
by the master equation when the independent thermal reservoirs are
considered. In section II, the time-dependent behaviour of quantum
discord and entanglement are analytically and numerically analyzed.
Compared with the case of noninteracting spin systems, the effects
of spin interactions on the decoherence of quantum discord are taken
into account. In section III, the thermal property of the quantum
discord is also investigated. Finally, a simple discussion concludes
the paper.

\section{The Markovian dynamics of quantum correlation}

To clearly describe the dynamics of quantum correlation, we use two
types of measures including the entanglement of formation and
quantum discord. For an arbitrary bipartite state $\rho_{AB}$, the
total correlations are expressed by quantum mutual information
\cite{Groisman05}
\begin{equation}
I(\rho_{AB})=\sum_{i=A,B}S(\rho_{i})-S(\rho_{AB}),
\end{equation}
where $\rho_{i}$ represents the reduced density matrix of subsystem
$i$ and $S(\rho)=-\mathrm{Tr}(\rho \log_{2}\rho)$ is the von Neumann
entropy. Henderson and Vedral proposed one measure of bipartite
classical correlation $C(\rho_{AB})$ based on a complete set of
local projectors $\{ \Pi_{B}^{k}\}$ on one subsystem $B$. After the
local measurements, the reduced state of subsystem $A$ can be
written by
\begin{equation}
\rho_{A}^{k}=\frac 1{p^{k}}\mathrm{Tr}_{B}[(\mathbf{1}_{A}\otimes
\Pi_{B}^{k})\rho_{AB}(\mathbf{1}_{A}\otimes \Pi_{B}^{k})].
\end{equation}
Here $p^{k}$ is the measurement probability for the $k-th$ local
projector and $\mathbf{1}_{A}$ is the identity operator on subsytem
$A$. When subsystem $B$ is described in the two-dimensional Hilbert
space $\{|0\rangle,|1\rangle \}$, a complete set of the local
measurements can be given by $\{ \Pi_{B}^{k}= V|k\rangle \langle
k|V^{\dag}, (k=0,1) \}$ where the arbitrary unitary operation
$V(\theta,\phi)$ for $\theta \in [0,\pi]$ and $\phi \in [0,2\pi]$ is
\begin{equation}
V(\theta,\phi)=\left(\begin{array}{cc}
            \cos \frac {\theta}2&\sin \frac {\theta}2 e^{-i\phi}\\
            \sin \frac {\theta}2 e^{i\phi}&-\cos \frac {\theta}2
            \end{array}\right).
\end{equation}
Then the classical correlation in the bipartite quantum state
$\rho_{AB}$ can be given by
\begin{equation}
C(\rho_{AB})=S(\rho_{A})-\sup_{\{ \Pi_{B}^{k}\}}\{S(\rho_{A|B}) \},
\end{equation}
where $S(\rho_{A|B})=\sum_{k=0,1}p^{k}S(\rho_{A}^{k})$ is the
conditional entropy of subsystem $A$ and $\sup \{S(\rho_{A|B})\}$
signifies the minimal value of the entropy with respect to a
complete set of local measurements $\{ \Pi_{B}^{k}\}$. Quantum
discord is simply defined by
\begin{equation}
Q(\rho_{AB})=I(\rho_{AB})-C(\rho_{AB}).
\end{equation}

In the following study, we draw on the master equation to analyze
the time-dependent behaviour of quantum discord in the quantum
system with the asymmetric anisotropic interaction. For a two-qubit
system, the total Hamiltonian is written by $H=H_{0}+H_{I}$ where
the intrinsic Hamiltonian of two qubits $H_{0}=\frac \omega{2}
(\sigma_A^z+\sigma_B^z)$ and the interacting Hamiltonian
$H_{I}=\frac
{J}{2}[\sum_{\alpha=x,y,z}\sigma_A^{\alpha}\sigma_{B}^{\alpha}+\vec{D}\cdot
(\vec{\sigma}_{A}\times \vec{\sigma}_{B})]$. To lay stress on the
effects of the interacting Hamiltonian on quantum discord, we assume
that $H_{0}\ll H_{I}$. Here $\sigma^{\alpha}$ is the pauli operator.
For the case of antiferromagnetic couplings $J>0$, the asymmetric
anisotropy $\vec{D}=D\vec{e}_{z}$ is considered.  It is reasonable
to assume that two qubits are coupled to its local thermal
reservoirs $E_A,E_B$ \cite{Yu04}. Suppose that the initial state at
$t=0$ is $\rho(0)=\rho_{AB}(0)\otimes
(|0_{E_A}0_{E_B}\rangle\langle0_{E_A}0_{E_B}|)$ where
$|0_{E_A}0_{E_B}\rangle$ denotes the vacuum state of the two local
reservoirs. The evolution of quantum state between qubit $A$ and $B$
is given by the master equation
\begin{equation}
\dot{\rho}_{AB}(t)=-i[H,\rho_{AB}]+\hat{L}(\rho_{AB}),
\end{equation}
where the Lindbald operator
\begin{align}
\hat{L}(\rho)=\sum_{i=A,B}
&(\bar{n}_i+1)\gamma_i(2\sigma_i^{-}\rho\sigma_i^{+}-\rho\sigma_i^{+}\sigma_i^{-}-\sigma_i^{+}\sigma_i^{-}\rho)&
\nonumber \\
&+\bar{n}_i\gamma_i(2\sigma_i^{+}\rho\sigma_i^{-}-\rho\sigma_i^{-}\sigma_i^{+}-\sigma_i^{-}\sigma_i^{+}\rho).&
\end{align}
Here $\bar{n}_i=\bar{n}$ is the mean number of the thermal reservoir
and $\gamma_i=\gamma$ signifies the rate of spontaneous emission for
each qubit. In fact, the temperatures of the thermal reservoirs can
be varied from zero values to small ones after the local operations.
At very low temperatures, the initial state is approximately the
ground state $\rho(0)$. The mean number $\bar{n}=\frac
{1}{\exp{(\omega_E/k_{B}T)}-1}$ could be a nonzero value where
$\omega_E$.

We adopt the numerical approach to calculate the density matrix at
any time. For numerical purposes, it is convenient to regard the
components of density matrix $\rho$ as being written as a single
column vector $\tilde{\rho}$ in Matlab. Then the pre-multiplication
of the operator $\sigma\rho$ can be written as
$spre(\sigma)\tilde{\rho}$ where the function $spre(\sigma)$ can be
given by the Kronecker product of the identity and matrix $\sigma$.
In the same way, the post-multiplication of the operator
$\rho\sigma$ can also be written as $spost(\sigma)\tilde{\rho}$.
Therefore, the master equation in Eq. (6) can also be expressed by
$\dot{\tilde{\rho}}_{AB}(t)=\mathcal{L}\tilde{\rho}_{AB}(0)$ where
the superoperator $\mathcal{L}$ is obtained by the
pre-multiplication and post-multiplication of the operators. Because
the superoperator $\mathcal{L}$ is time-independent and is small
enough to be diagonalized numerically, the $i-th$ component of
$\tilde{\rho}(t)$ can be calculated by $\rho_i(t)=\sum_j a_{ij}\exp
s_{j}t$. Here $s_j$ is the eigenvalue of $\mathcal{L}$ and satisfies
$\mathcal{L}_{ik}=\sum_j U_{ij}s_{j}U^{-1}_{jk}$. The coefficients
$a_{ij}=U_{ij}\sum_k U^{-1}_{jk}\tilde{\rho}_k(0)$. The numerical
calculation of $a_{ij}$ and $s_j$ is completed by the program of
exact diagonalization of the superoperator $\mathcal{L}$.

From the point of view of the practice, the initial state
$\rho_{AB}(0)$ between qubit $A$ and $B$ is usually chosen to be the
ground state which is expanded in the product space of two qubits
\begin{equation}
|\psi_{g}\rangle=\frac 1{\sqrt2}(|01\rangle+e^{i\alpha}|10\rangle),
\end{equation}
where $e^{i\alpha}=\frac {1+iD}{\sqrt{1+D^2}}$. It is clearly seen
that the ground state is just one maximally entangled Bell state.
The general expression of quantum state $\rho_{AB}(t)$ satisfies
\begin{equation}
\rho_{AB}(t)=\left(\begin{array}{cccc}
            u&0&0&0\\
            0&x&z&0\\
            0&z^{\ast}&y&0\\
            0&0&0&v
            \end{array}\right).
\end{equation}
According to the result of \cite{Mazhar10}, the analytical solution
of quantum correlation can be given by
\begin{equation}
Q(\rho_{AB})=S(\rho_{B})-S(\rho_{AB})+\min{\{S_0,S_1\}}.
\end{equation}
The entropy is $S_{0}=-(u\log_2\frac {u}{u+y}+y\log_2\frac
{y}{u+y}+x\log_2\frac {x}{x+v}+v\log_2\frac {v}{x+v})$ and
$S_{1}=-\sum_{q=0,1}\frac {1+(-1)^q \theta_1}{2}\log_2 \frac
{1+(-1)^q \theta_1}{2}$ where $\theta_1=\sqrt{(u+x-y-v)^2+4|z|^2}$.
Meanwhile, the entanglement of formation can be used to evaluate the
quantum correlation, According to \cite{Wang02,Arnesen01,Bayat05},
the value of the entanglement is obtained by
\begin{equation}
E(\rho_{AB})=2\max{\{ |z|-\sqrt{uv},0 \}}.
\end{equation}
Through the numerical approach, the decoherence of quantum
correlation can be given by Fig. 1(a) and 1(b). If qubits $A$ and
$B$ are in the ground state, the maximally entangled ground state
has the maximal value of quantum correlation,
$E(\rho_{AB})=Q(\rho_{AB})=1$ at the time $t=0$. In Fig. 1(a), it is
clearly shown that the values of quantum discord $Q(\rho_{AB})$ are
decreased and infinitely close to zero with time. The nonvanishing
phenomenon of quantum discord is different from the sudden
disappearance of the entanglement. In the vicinity of
$E(\rho_{AB})=0$, the nonzero values of the discord can also
quantify the nonclassical correlation for separable mixed states
which make possible some tasks of quantum computers. However, it is
noted that the decay of quantum discord is very similar to that of
the entanglement before the time of the vanishing of entanglement.
This aspect demonstrates that quantum discord can describe the
nonlocal coherence like the entanglement. It is also valuable to
investigate the case where qubits are initially in a separable state
$\psi_{AB}(0)=|10\rangle$. From Fig. 1(b), we find that quantum
discord and entanglement can be generated and then vanish in time.
The behaviour arises from the coupling between qubit $A$ and $B$.

For a special case of $J=0$, the evolution of quantum state
$\rho_{AB}(t)$ can be described by a completely positive
trace-preserving map \cite{Aolita08}. For a general two-qubit state
$\rho_{AB}(0)=\sum_{kl,mn}a_{mn,kl}|k\rangle_{A}\langle m|\otimes
|l\rangle_B \langle n|$, the evolved state in time can be written by
$\rho_{AB}(t)=\sum_{kl,mn}\sum_{j=0}^{3}\sum_{j'=0}^{3}a_{mn,kl}(K_{Aj}|k\rangle_A\langle
m|K^{\dag}_{Aj})\otimes(K_{Bj'}|l\rangle_B\langle n|K^{\dag}_{Bj'})
$ where the Kraus operators $K_{i0}=\sqrt{\frac
{\bar{n}+1}{2\bar{n}+1}}(|0\rangle_i\langle
0|+\sqrt{1-p}|1\rangle_i\langle 1|)$, $K_{i1}=\sqrt{\frac
{(\bar{n}+1)p}{2\bar{n}+1}}|0\rangle_i\langle 1|$,
$K_{i2}=\sqrt{\frac
{\bar{n}}{2\bar{n}+1}}(\sqrt{1-p}|0\rangle_i\langle
0|+|1\rangle_i\langle 1|)$ and $K_{i3}=\sqrt{\frac
{\bar{n}p}{2\bar{n}+1}}|1\rangle_i\langle 0|$. Here $|0(1)\rangle_i$
is the ground(excited) state of qubits $i=A,B$ satisfying
$\sigma_{i}^{z}|0(1)\rangle_i=\mp|0(1)\rangle_i$ and
$p(t)=1-e^{\frac {\gamma(2\bar{n}+1)t}{2}}$ means the probability of
the atom exchanging a quantum with the reservoir. If the initial
state is $|\psi\rangle_{g}$, the expression of $\rho_{AB}(t)$ is
similar to Eq. (9),
\begin{align}
u=&\frac {\bar{n}p}{2\bar{n}+1}[1-\frac
{(\bar{n}+1)p}{2\bar{n}+1}],v=\frac
{(\bar{n}+1)p}{2\bar{n}+1}(1-\frac
{\bar{n}p}{2\bar{n}+1}),&\nonumber\\
x=&\frac 12[\frac {\bar{n}(\bar{n}+1)p^2}{(2\bar{n}+1)^2}+(1-\frac
{\bar{n}p}{2\bar{n}+1})(1-\frac {(\bar{n}+1)p}{2\bar{n}+1})],&
\nonumber \\
z=&\frac 12(1-p),y=1-u-x-v.&
\end{align}
Therefore, the angle $\theta_{1}$ of the entropy $S_{1}$ in Eq. (10)
is analytically expressed by $\theta_{1}=\sqrt{\frac
{p^2}{(2\bar{n}+1)^2}+(1-p)^2}$. From the above analysis, it is seen
that quantum discord is independent on the asymmetric anisotropy $D$
if the initial state is the ground state $|\psi_{g}\rangle$. In Fig.
2, two cases of the decoherence of quantum discord are shown when
the interactions between two qubits are $J=0$ and $J=1$
respectively. It is found that the values of quantum discord rapidly
decline to very small ones under the impact of the interaction
$J\neq 0$. This manifests that the dramatic loss of nonclassical
correlation happens when there is the strong interaction between
qubits in the system.

\section{Thermal decoherence of quantum discord}

In general, states of qubits for quantum information processing are
usually influenced by thermal temperatures. Under the effect of the
temperatures, the thermal decoherence is unavoidable. Therefore, it
is of value to study the properties of quantum discord for the
thermal equilibrium state. In the spin model, the mixed state
$\rho(T)$ at any equilibrium temperature $T$ is expressed by
\begin{equation}
\rho(T)=\frac 1{Z}\sum_{j}|\psi_{j}\rangle\langle \psi_{j}|
\exp(-\epsilon_j/kT),
\end{equation}
where the partition function $Z=\sum_{j}\exp(-\epsilon_j/kT)$ and
$|\psi_{j}\rangle$, $\epsilon_j$ are the $j-th$ eigenstate of the
Hamiltonian $H$ and corresponding eigenvalue respectively. The exact
diagonal expression of $H$ is given by a set of eigenstates and
corresponding eigenvalues,
\begin{align}
\epsilon_{1,2}=&J(-\frac 12\pm |\eta|),|\psi_{1,2}\rangle=\frac
1{\sqrt2}(|01\rangle\pm\frac
{\eta}{|\eta|}|10\rangle);&\nonumber \\
\epsilon_{3}=&\frac J2-\omega,|\psi_{3}\rangle=|00\rangle;
\epsilon_{4}=\frac J2+\omega,|\psi_{4}\rangle=|11\rangle.&
\end{align}
Here the parameter $\eta=1+iD$. Because the thermal state has the
same form in Eq. (9), the calculation of quantum discord is also
solved by Eq. (10). The thermal properties of quantum discord are
shown in Fig. 3. It is found that the asymmetric anisotropy $D$ can
increase the values of quantum discord at any temperature $T$. This
is the reason that the asymmetric anisotropy along the $z$ direction
induces the quantum fluctuation in the $XY$ plane which can enhance
the quantum correlation. The decay of quantum discord occurs when
the temperatures $T$ are increased. For finite temperatures, the
values of quantum discord are always nonzero. The result coincides
with that of \cite{Werlang10}. As is well known, the thermal
entanglement is gradually decreased to zero with the increase of the
temperatures \cite{Arnesen01}. In the context, it is the fact that
the robustness of quantum discord against the decoherence is helpful
for the realization of quantum computing.

\section{Discussion}

Using quantum discord, we mainly investigate the decoherence of
quantum correlation for two interacting qubits coupled to
independent reservoirs. Compared with the case of noninteracting
qubits, quantum correlation of the states in the interacting qubits
system decays very rapidly. The values of quantum discord decrease
asymptotically in time. This nonvanishing phenomenon is apparently
different from the sudden change of the entanglement. It is found
that the evolution behaviour of quantum discord is very similar to
that of the entanglement before the sudden disappearance of the
entanglement. This demonstrates that quantum discord for inseparable
states can describe the nonlocal coherence like the entanglement.
Moreover, after the sudden death of the entanglement, the nonzero
values of quantum discord also manifest the existence of quantum
correlation for separable mixed states. The thermal decoherence from
the temperature is also considered. It is shown that the
nonvanishing quantum discord at any finite temperature can be
increased by the asymmetric anisotropy. Quantum discord also
declines asymptotically with the temperature. Therefore, quantum
discord is a general quantifier for nonclassical correlation.

\section{Acknowledgements}

The work was supported by the Research Program of Natural Science
for Colleges and Universities in Jiangsu Province Grant No.
09KJB140009 and the National Natural Science Foundation Grant No.
10904104.

\newpage

{\Large Figure caption}

Figure 1

As two different quantifiers of quantum correlation, the quantum
discord (solid line) and entanglement (dashed line) are numerically
calculated and plotted as a function of the decoherence time. The
mean number of the thermal reservoir is chosen to be $\bar{n}=1$ and
the rate of spontaneous emission is $\gamma=0.1$. The parameters are
chosen to be $J=1$, $D=0.2$ and $\omega=0.1$. (a). The qubits are
initially in the ground state $|\psi_{AB}(0)\rangle=|\psi_g\rangle$;
(b). The initial state of the qubits is a separable one
$|\psi_{AB}(0)\rangle=|10\rangle$.

Figure 2

The decoherence of quantum discord for the two-qubit system is
illustrated under the independent reservoirs. The solid line denotes
the case of two interacting qubits in the condition of $J=1$,
$D=0.2$ and $\omega=0.1$. The dashed line signifies the other case
of two noninteracting qubits for $J=0$. The mean number of the
thermal reservoir is chosen to be $\bar{n}=1$ and the rate of
spontaneous emission is $\gamma=0.1$.

Figure 3

The quantum discord at the thermal equilibrium state for $J=1$ and
$\omega=0.1$ is shown with the changes of the temperature $T$ and
asymmetric anisotropic interaction $D$.

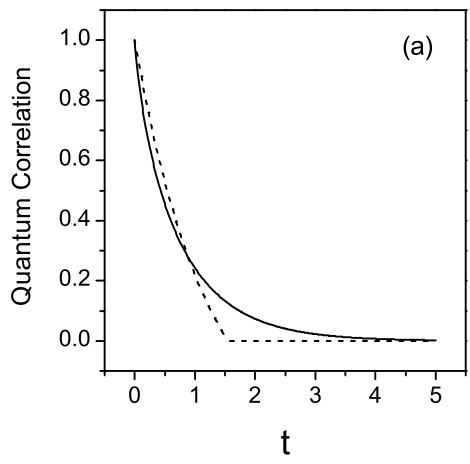
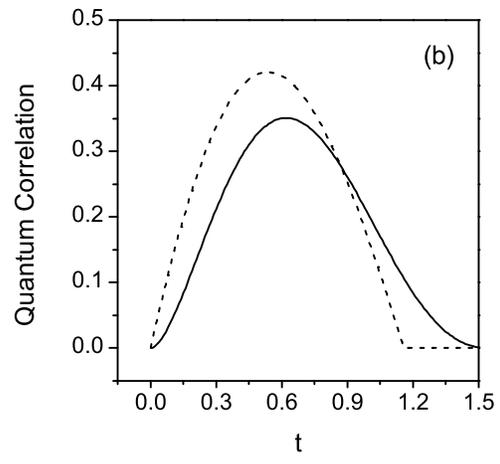
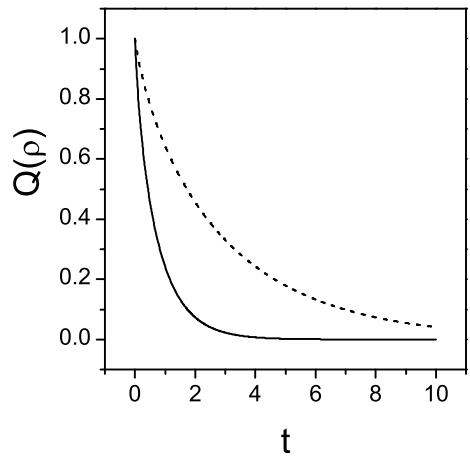
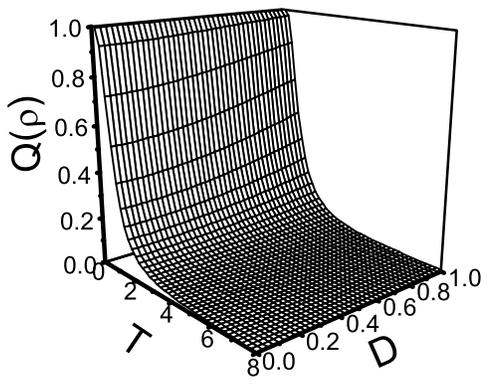
\end{document}